# DEM simulation of granular segregation in two-compartment system under zero gravity*

Wenguang Wan(王文广)[1], Zhigang Zhou(周志刚)[1], Jin Zong(宗谨)[1], and MeiyingHou(厚美瑛)[1,2†]

[1] *Key Laboratory of Soft Matter Physics, Beijing National Laboratory for Condense Matter Physics, Institute of Physics, Chinese Academy of Sciences, Beijing 100190, China*
[2]*College of Physics, Chinese Academic of Sciences University*

In this paper, granular segregation in a two-compartment cell in zero gravity is studied numerically by DEM simulation. In the simulation using a virtual window method we find a non-monotonic flux, a function which governs the segregation. A parameter $\varepsilon$ is used to quantify the segregation. The effect of three parameters: the total number of particles $N$, the excitation strength $\Gamma$, and the position of the window coupling the two compartments, on the segregation $\varepsilon$ and the waiting time $\tau$ are investigated. It is found that the segregation observed in zero gravity exists and does not depend on the excitation strength $\Gamma$. The waiting time $\tau$, however, depends strongly on $\Gamma$: Higher the $\Gamma$, lower the waiting time $\tau$. The simulation results are important in guiding the SJ-10 satellite microgravity experiments.

## 1. Introduction

Granular materials' extremely rich dynamical behaviors have attracted attentions of physicists of different fields in recent years[1-3]. Examples are the heap formation of a granular bed[4-6] and the size segregation of a granular system with grains of various sizes under vertical vibration known as the Brazil Nut effects[7-8]. In these phenomena, energy is being injected continuously into the system by the oscillating boundaries and propagated into the bulk by the inelastic collisions of the grains. A steady state of the whole system is reached when the dissipation of the system is balanced by the input of the energy. Since both the energy input and dissipation depends crucially on the configurations of the system, many intriguing steady states and even oscillatory states can be created.

Borrowing concepts from molecular gas system, we treat low density granular system as a granular gas. The granular gas systems reach a steady state when input and loss of energy are balanced. They are not in thermal equilibrium and the laws of thermodynamics for molecular gases do not apply for these systems. For example, the thermodynamically impossible phenomenon such as the Maxwell's demon[9,10] has been observed and successfully explained. In such an experiment, granular gas confined in a compartmentalized system can be induced to segregate into one of the compartments by lowering the vibration amplitude of the system. In this latter case, a decrease of the configurational entropy of the system takes place spontaneously; as if the second law of thermodynamics is violated. In fact, other similar intriguing segregation[11] and ratchet effects[12] have also been reported in compartmentalized granular gases.

* Project supported by the "Strategic Priority Research Program-SJ-10" of the Chinese Academy of Sciences (Grant No.XDA04020200), the National Natural Science Foundation of China (Grant Nos. 11274354, 11474326), and the Special Fund for Earthquake Research of China (Grant No. 201208011).
† Corresponding author. E-mail: mayhou@iphy.ac.cn

Granular Maxwell's Demon phenomenon has been studied in simulation[13-23], theoretical modeling[24-29] and by experiment[30-40] in recent years extensively. The segregation phenomenon relies on the existence of a non-monotonic flux, which determines the number of particles per unit time flows between the two compartments[10]. The flux function is derived from the equation of gas state and herefore depends on the gravity. If applicable, the phenomenon can be used to transport granular materials in space[41,42]. In this work, we perform simulation study to find the condition for possible segregation in two-compartment granular gas system in an environment of zero gravity. We find (1) density gradient exists in the cell along excitation direction that particles gather near the end away from the vibration wall; (2) segregation among the two compartments exists even for very small excitation acceleration $\Gamma$ as long as number of particles $N$ exceeds a critical number and the waiting time is long enough; (3) segregation quantified by an asymmetry parameter $\varepsilon$ depends weakly upon the position of the opening connecting the two compartments, but different from the situation when in gravity, $\varepsilon$ is independent of $\Gamma$ when in zero gravity; (4) the waiting time $\tau$ is shorter when the acceleration $\Gamma$ is higher, but reaches a minimum when $\Gamma$ is large enough.

## 2. Model

The simulation is based on discrete element method, in which each particle is treated as a discrete element[43,44]. The motion of each particle obeys the Newton's Second Law. The interaction between them is considered if and only if two particles collide. Taking the rotation into consideration, the particles' equations of motion are as follows:

$$m_i \ddot{\vec{r}}_i = \sum_{j,j\neq i} \vec{F}_{ij}^C + \vec{F}_i^O , \tag{1}$$

$$I_i \dot{\vec{\omega}}_i = \sum_{j,j\neq i} \vec{R}_i \times \vec{F}_{ij}^C + \vec{M}_i^O , \tag{2}$$

where $m_i$, $I_i$, $\vec{r}_i$ and $\vec{\omega}_i$ are the mass, the moment of inertia, the displacement and the angular velocity of particle $i$, respectively. $\vec{F}_{ij}^C$ is the contact force that the particle $i$ is applied by the particle $j$. $\vec{R}_i$ is the radius vector of particle $i$. $\vec{F}_i^O$ and $\vec{M}_i^O$ are other forces and torques applied to the particle $i$, such as the gravity force. Generally, $\vec{F}_{ij}^C$ can be decomposed into two parts: a normal component and a tangential one,

$$\vec{F}_{ij}^C = \vec{F}_{ij}^n + \vec{F}_{ij}^t . \tag{3}$$

The normal and the tangential contact forces between two particles are handled separately.

Two particles interact with each other only during contact. The overlap in the normal direction between particles $i$ and $j$ with radii $R_i$ and $R_j$, respectively, is

$$\delta = \left(R_i + R_j\right) - \left(\vec{r}_i - \vec{r}_j\right) \bullet \vec{n}, \tag{4}$$

Where $\delta$ is a positive number when the two particles interact, $\vec{n}$ is the unit vector of the relative displacement $\vec{n} = \left(\vec{r}_i - \vec{r}_j\right) / \left|\vec{r}_i - \vec{r}_j\right|$. Considering the dissipative effect, the equation of the normal contact force is

$$F_{ij}^n = k_n \delta + \gamma_n v_n, \tag{5}$$

where $k_n$ and $\gamma_n$ are the spring stiffness and the dissipative coefficient, respectively, and $v_n$ is the normal component of the relative velocity $\vec{v}_{ij} = \vec{v}_i - \vec{v}_j$. This is called Linear Spring-Dashpot (LSD) model. Linear model approximation is applicable in our case for the dominant short-contact two-body collisions and the small deformation $\delta$. Comparing to Hertz model, linear Hooke's law has the advantage that the contact duration does not depend on the impact velocity of the particles, and it allows us to solve analytically the equation of motion.

The tangential contact force is determined by the tangential component of the relative velocity and the normal contact force

$$F_{ij}^t = \min\left(\gamma_t v_{ij}^t, \mu F_{ij}^n\right), \tag{6}$$

Where $\gamma_t$ is a numerical constant, $\mu$ is the friction coefficient and $v_{ij}^t$ is the tangential component of $\vec{v}_{ij}$. Fig. 1 shows the mechanism of the contact model in this work.

According to the LSD model, the collision process between two particles acts like a damped harmonic oscillator. We can get the contact duration $t_C$, half of the period of the damped harmonic oscillator, analytically,

$$t_c = \pi / \sqrt{k_n / m_{ij} - \gamma_n^2 / 4m_{ij}^2}, \tag{7}$$

Where $m_{ij}$ is the reduced mass $m_{ij} = m_i m_j / \left(m_i + m_j\right)$. Using the oscillation half period, we obtain the coefficient of restitution by the ratio of the final velocity and the initial velocity,

$$e = v_{final} / v_0 = \exp\left(-\gamma_n t_c / 2m_{ij}\right). \tag{8}$$

By solving Eq.(8), we can get the dissipative coefficient $\gamma_n$ via the restitution coefficient $e$,

$$\gamma_n = -2\sqrt{k_n m_{ij}} \ln\left(e\right) / \sqrt{\ln^2\left(e\right) + \pi^2}. \tag{9}$$

Using LSD model, the coefficient of restitution is a constant independent of the relative velocity of two particles during collision. Additionally, in order to guarantee the simulation's accuracy, the time duration per step $\Delta t$ is required to be much less than $t_C$, i.e. $\Delta t \ll t_c$. In the simulation we take $\Delta t \approx t_C / 10$.

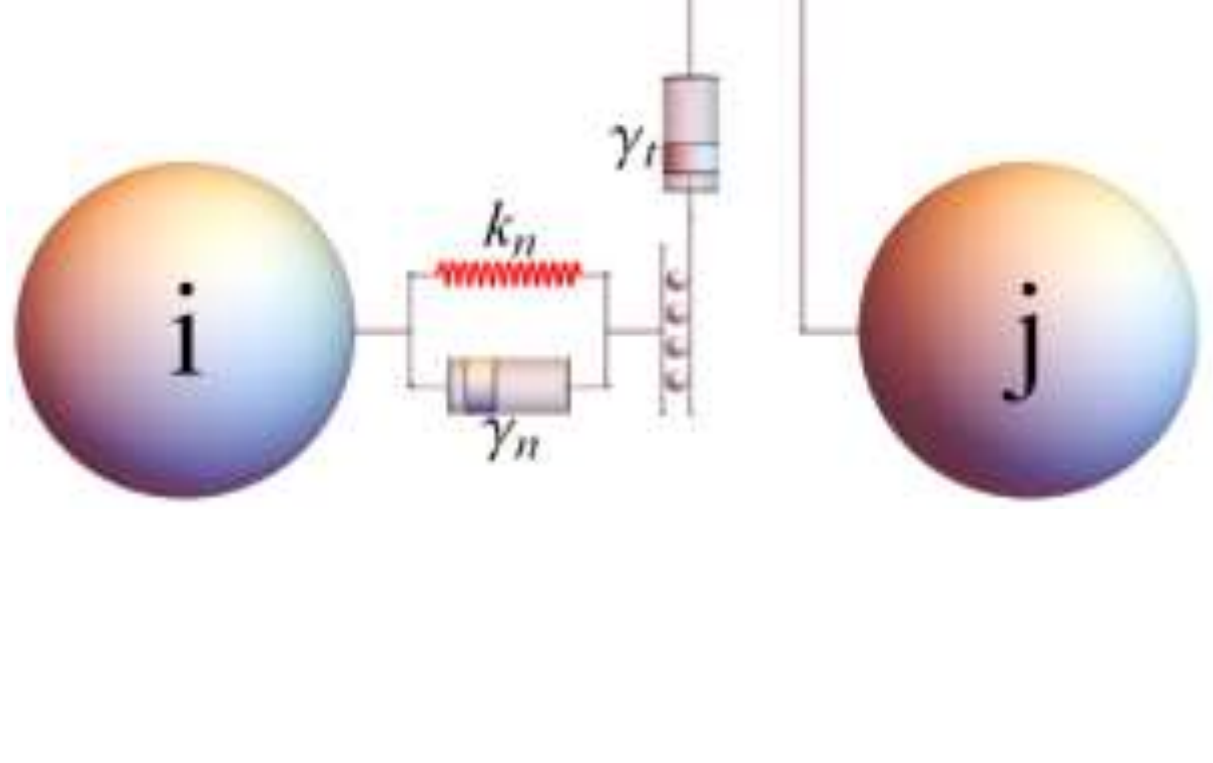

Fig. 1. The mechanism of the contact model.

## 3. Setup

### 3.1. Experimental Setup

Fig. 2(a) is a photo of the two-compartment sample cell used in the experiment of the recoverable satellite “SJ-10”. Grains in the cell are driven by four independently moving pistons, which are used as the end walls of the two compartments as plotted in Fig. 2 (b). The two compartments are connected by a window of 12mm wide. The total length of the cell is 100 mm, as seen in Fig. 2(b), and the individual piston size is 25 mm × 25mm. The four pistons are driven by two linear motors. Each motor controls either left- or right-end two pistons, so that the two pistons at the same side vibrate synchronously. The piston’s position, vibration frequency and amplitude are pre-programmed. In order to prevent the electrostatic effect, metallic particles are used and the cell is made of metal and has been grounded. The fused silica window plates are coated by conductive thin films. Particles are enclosed between the left and right pistons. The volumes of the two cells can then be controlled by the positions of the pistons.

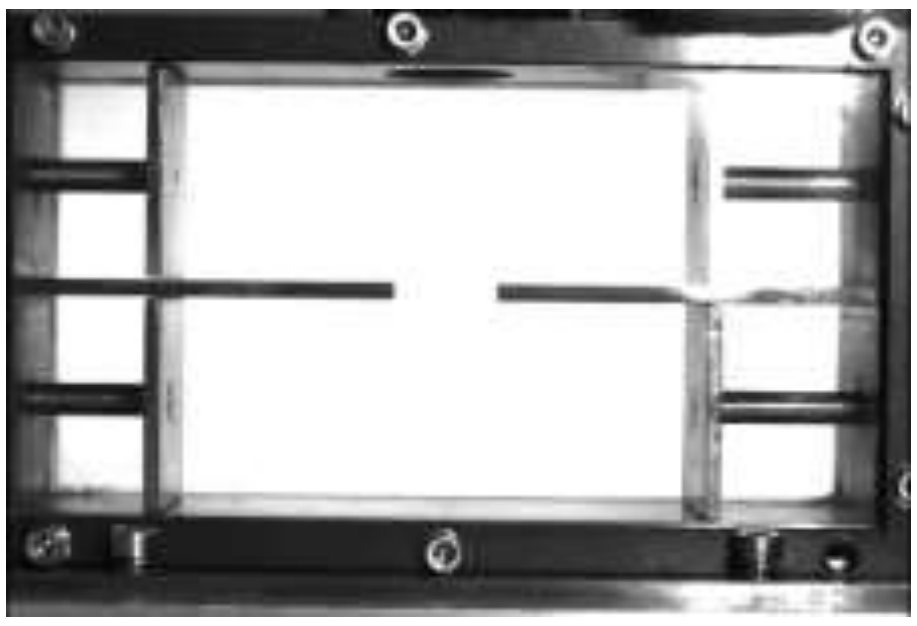

(a)

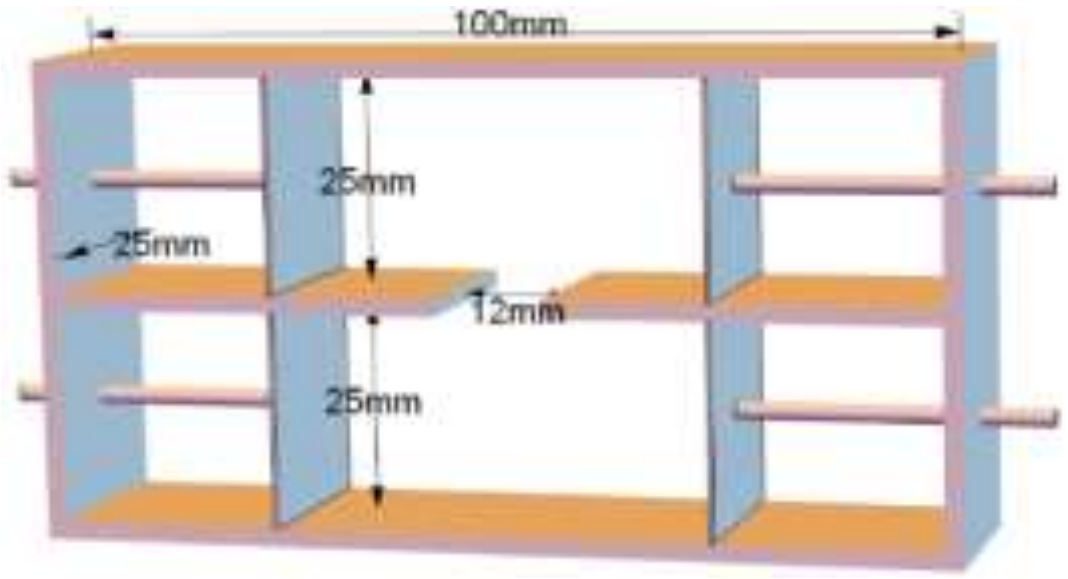

(b)

Fig. 2. (a) A photograph of the two-compartment sample cell used in the experiment of the recoverable satellite “SJ-10”; (b) sketch of the setup.

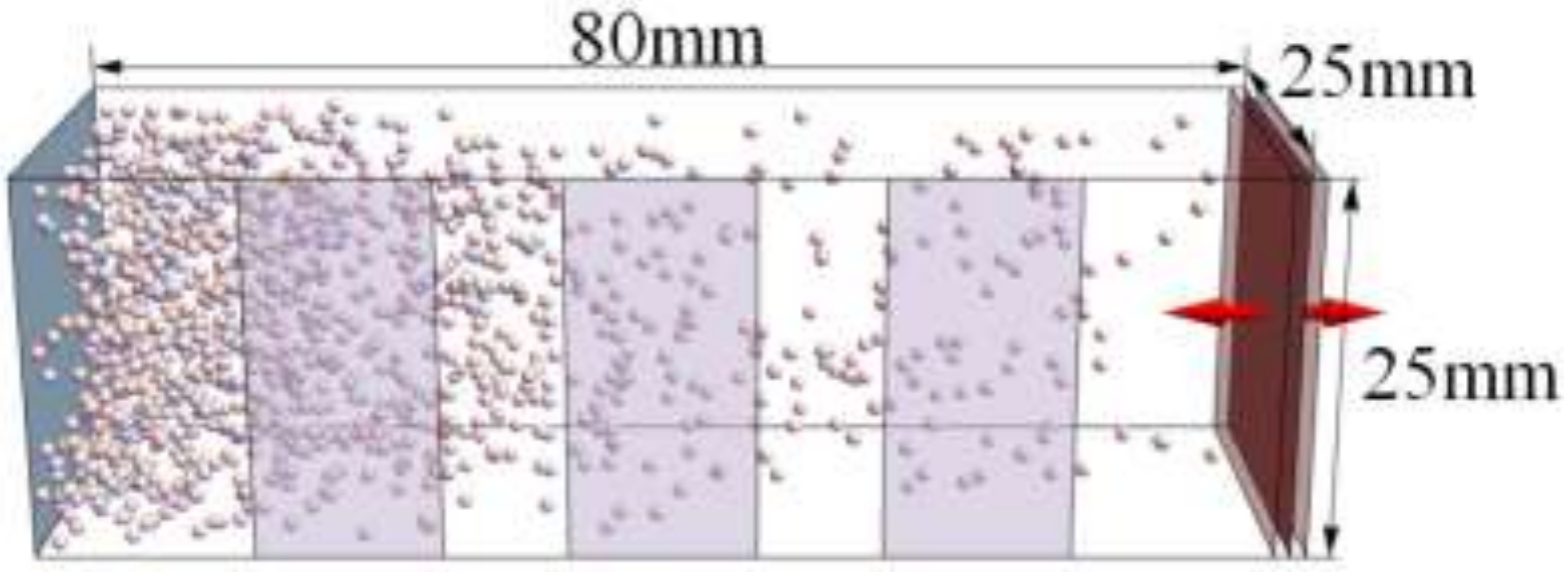

Fig. 3. Schematic of the virtual-window single cell. Particles are excited by the right-end moving wall. The virtual window at three different locations are indicated by rectangular shades at the side wall.

### 3.2. Numerical Setup

#### 3.2.1. Double-cell

The numerical parameters of the granular cell system are taken same as that in the experiments shown in Fig. 2(b). In the simulation, the particle radius $R$ is 0.5mm. The density $\rho$ of particles is 4500kg/m$^3$. The coefficient of restitution $e$ for grain-grain collision is taken as 0.8. The coefficient of frication $\mu$ is 0.3. The total number of particles in the connected two-compartment cell is taken as $N$. The number in one (say upper one) cell is $N_1$, and the number in the other (lower) cell is $N_2 = N - N_1$. To start with, we put particles evenly in each compartment, $N_1 = N_2 = N/2$. The initial velocity is arbitrarily set to be 0.1m/s with random direction.

### 3.2.2. Single-cell

Simulations of particles in single cell are performed to investigate in zero gravity the particle distribution along the vibration direction. With gravity, using the equation of gas state, J. Eggers[10] sets up a flux function to model the segregation phenomenon. In space under microgravity we need to know the profile of the particle distribution to set up a flux function in order to find the segregation condition. To know the number of particles flow instantaneously through the window per unit time, we count the number of particle-wall collisions at the virtual window of the cell[21]. The size of the single-cell is 80mm×25mm×25mm. Three rectangular shades along the side wall are shown in Fig. 3 to indicate the virtual windows at different positions. We record how many times particles collide with a given virtual window at a time interval for a sufficiently long time. The flux profile can then be given by the counts of collisions per unit time across a given virtual window.

## 4. Results

### 4.1. Distribution of particles in single cell

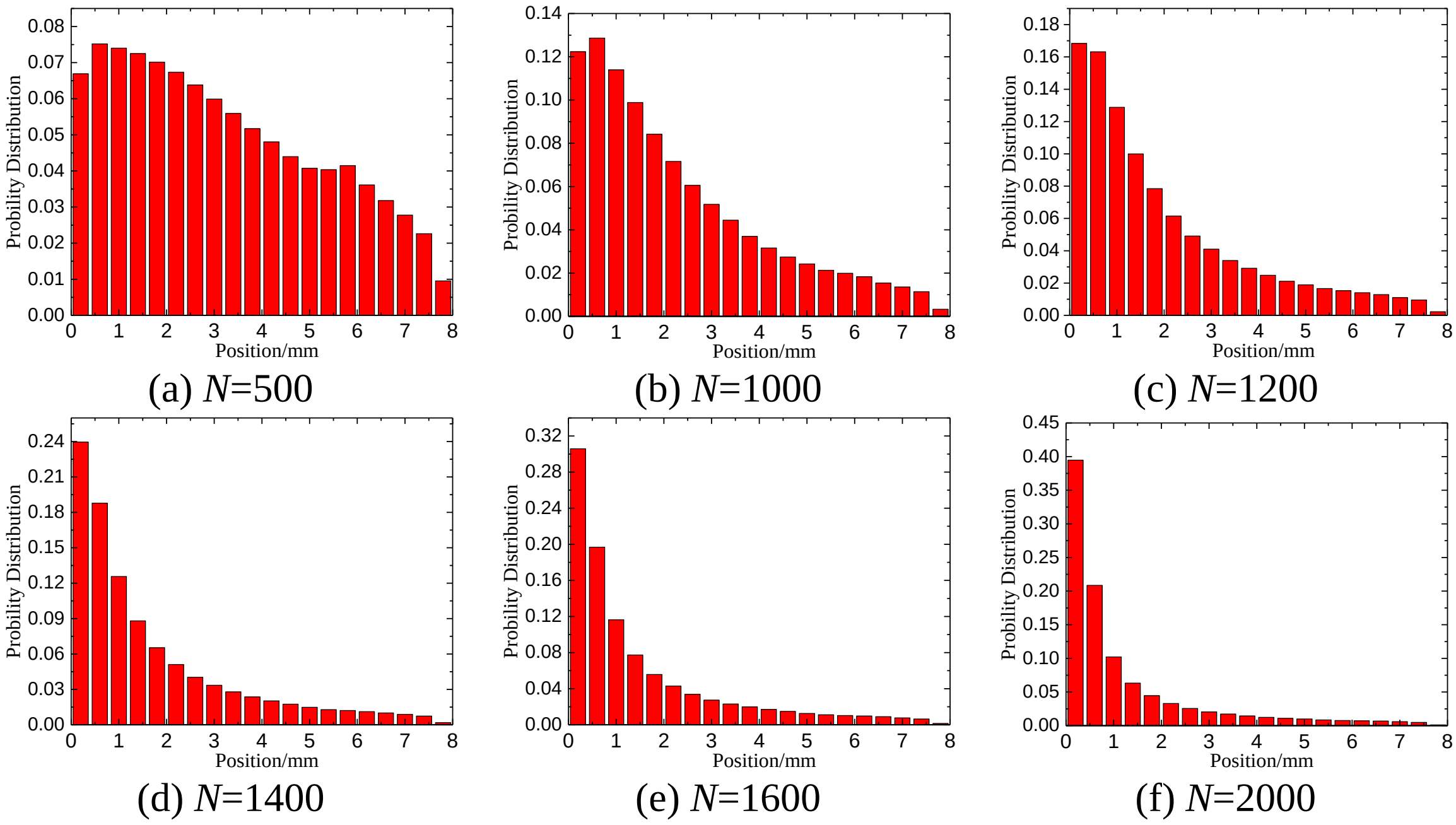

Fig. 4. Density distribution of particles in single cell with different numbers of particles.

Under gravity in J. Eggers' model[10] the equation of gas state is conveniently used for the density distribution of particles along the vibration direction. In microgravity the particle distribution is determined by the geometry of the cell and the wall excitation condition. We, therefore, perform a numerical study in a same cell geometry configuration and wall excitation condition to obtain the zero-gravity particle distribution. One set of the simulation results at condition that frequency $f_r$ =10Hz and shaking amplitude $A_r$ = 3mm, is shown in Fig. 4. Dividing the cell into 20 equal parts along the vibration direction, particle counts in each division are averaged for different total number $N$ . Except for situation in Fig. 4(a) when $N$=500, particle distributions in steady state for N greater than 1000 show exponential distribution along x-axis, which means similar distribution as the situation with gravity is achievable. For $N$ is less than 1000, the distribution is relatively homogeneous. With greater number $N$ , particles gather denser to the cool end. When the number $N$ is 2000, nearly 40 percent of particles cluster in 5 percent of the total volume.

### 4.2. Flux function obtained in virtual-window single-cell simulation

Finding flux function is necessary to understand the occurrence of the segregation. The segregation is governed by the following equation:

$$\frac{dN_i(t)}{dt} = -F\left(N_i(t)\right) + F\left(N - N_i(t)\right), \quad i = 1,2, \tag{10}$$

where the $N_1(t)$ and $N_2(t)$ are the counts of particles in the two compartments at time t, and $F(N_i)$ is the number of particles per unit time at time t that flow from a compartment containing $N_i$ particles to another compartment. Once knowing the function $F(N_i)$, $N_i(t)$ can be obtained from Eq. (10).

In order to characterize the segregation of particles in the two cells, a dimensionless parameter $\varepsilon_i$ is introduced:

$$\varepsilon_i = \left(N_i - N/2\right)/N, \quad i = 1,2. \tag{11}$$

Fig. 5 shows a plot of simulation result how $\varepsilon_i$ changes with time. It takes some time for the population $N_i(t)$ to reach a steady state. We call this time as the waiting time $\tau$ . At any time $\varepsilon_1(t) = -\varepsilon_2(t)$ , and the absolute value of $\varepsilon_i$ at the steady state is indicated as $\varepsilon$ . $\varepsilon = 0$ means particles are equally populated in the two cells, and $|\varepsilon| = 0.5$ means all the particles gather in one compartment, i.e., fully segregated.

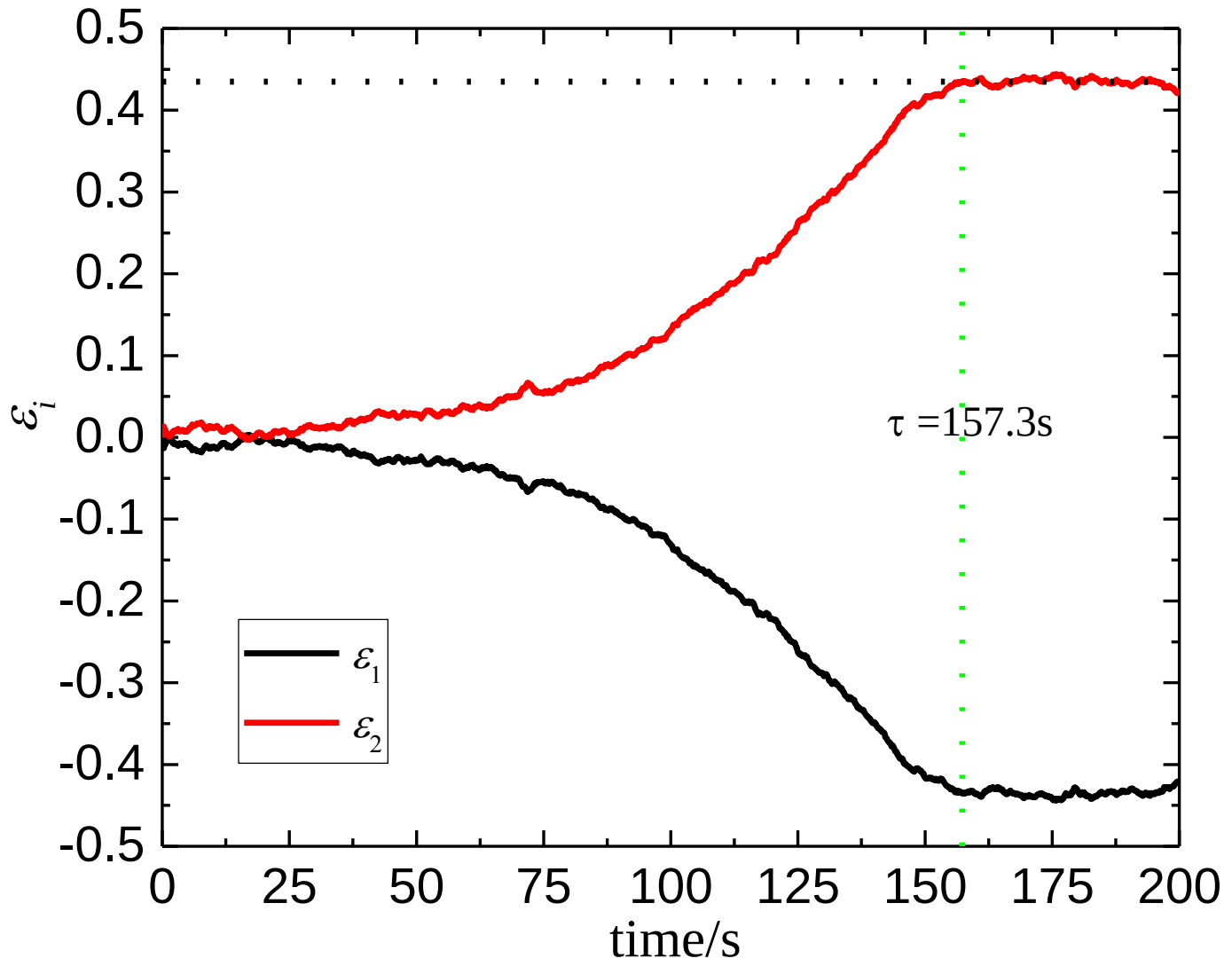

Fig. 5. One example of simulation result of $\varepsilon_i(t)$

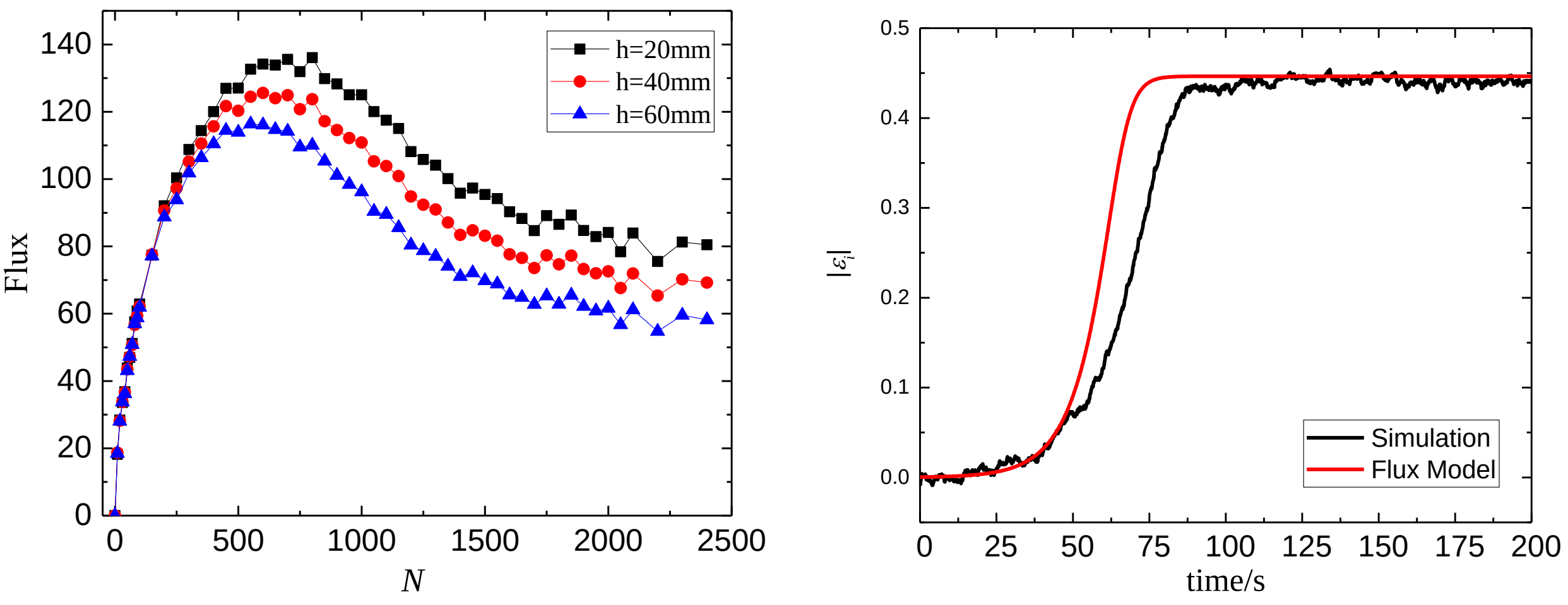

(a) (b)

Fig. 6. (a)Flux as a function of the number $N$ ; (b) Comparison between the theoretical result from the flux model and the simulation result.

The flux function obtained by the virtual-window single-cell simulation method[21] is shown in Fig. 6(a) for three locations of the window. A non-monotonic function $F(N_i)$ guaranteed the occurrence of the segregation. A weakly position dependence of the flux function shown in Fig. 6(a) tells us that flux value is greater when the window position is moving away from the moving wall. $F(N_i)$ is solved numerically. Numerically solving the Eq.(10), from $N_i(t)$ we get the value $|\varepsilon_i(t)|$ and compare it with the simulation results, as shown in Fig. 6(b). The steady values $\varepsilon$ obtained by the flux model and the simulation are in good agreement. In the flux model F is only a function of $N_i$. The simulation results, however, depend not only on $N_i$, but also on the initial velocity distribution of particles in each compartment. Therefore on average the waiting time shall be comparable between simulation and modeling. For a single run of simulation as shown in Fig. 6(b), discrepancy in time is expected depending on the chosen initial velocities of particles for the simulation run.

### 4.3. The effect of total number $N$ on the segregation $\varepsilon$

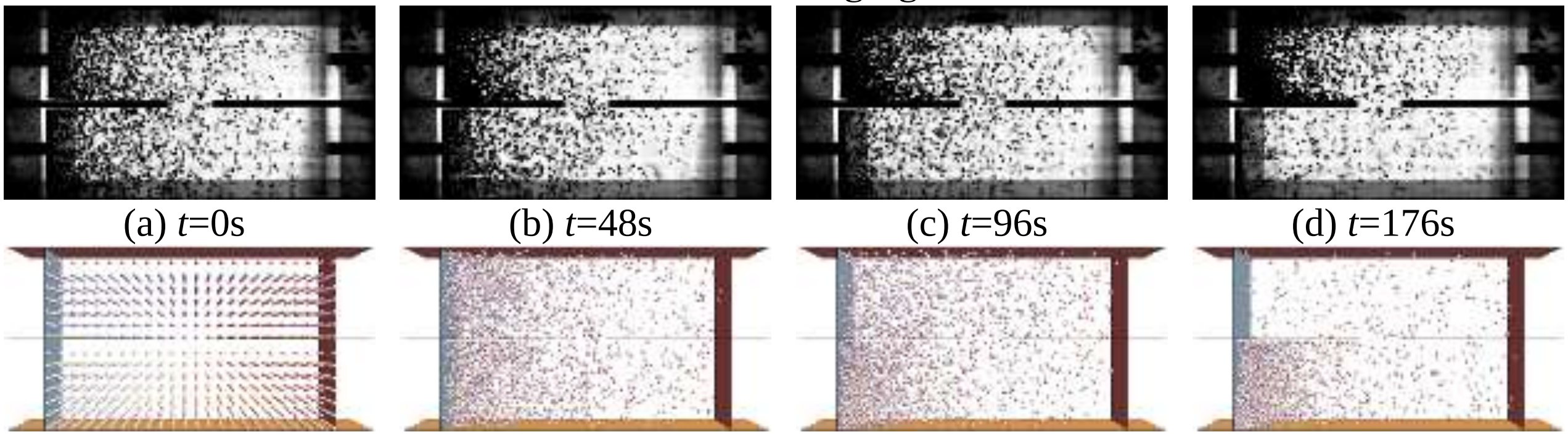

(a) *t*=0s (b) *t*=48s (c) *t*=96s (d) *t*=176s

(e) *t*=0s (f) *t*=50s (g) *t*=100s (h) *t*=200s

Fig. 7. (a)-(d) Snapshots captured in the experiments at different time; (e)-(h) Snapshots in the simulations taken at similar time.

In the SJ-10 experiment, the segregation is observed at total particle number N= 2400, cell length = 80 mm, and with right-end shaking at frequency f = 7 Hz, amplitude A=2 mm, (equivalently acceleration $\Gamma$ = 0.38 $g$ ). The vibration intensity is characterized by the acceleration $\Gamma = A_r\left(2\pi f_r\right)^2 / g$, where $g$ is the acceleration of gravity. In the simulation, we set amplitude A to be fixed and change frequency of the right piston to change the shaking acceleration Γ. Fig. 7(a)-4(d) are snapshots captured in the experiment at four different time. It shows the distribution segregation is fully developed at the time t=176 s.

Using similar parameters, a simulation is performed and a comparison shows consistent results as seen in Fig. 7(e)-4(h). Initially particles are distributed equally in the two cells. Segregation appears after about 100 seconds, and is fully developed 200 seconds afterwards.

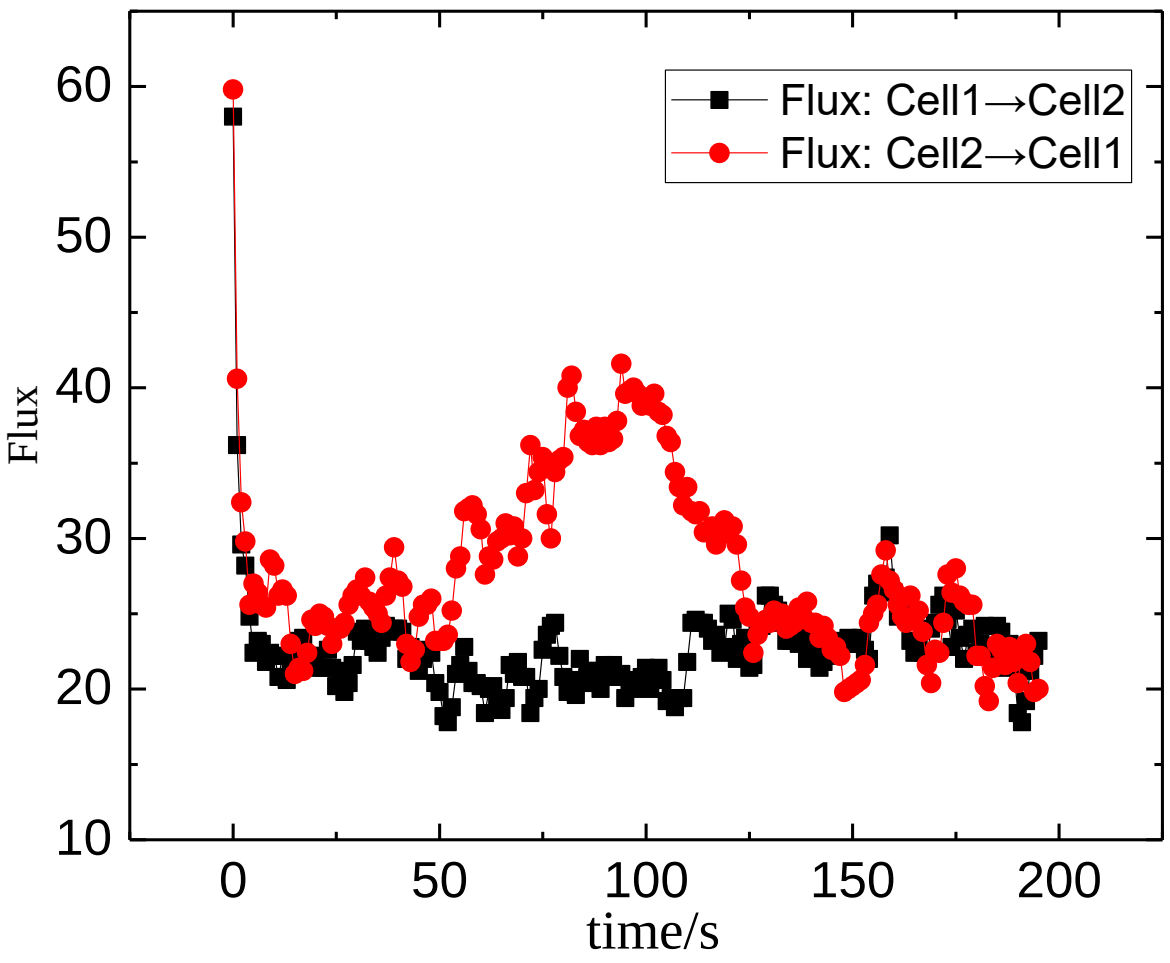

Fig. 8. Flux as a function of time.

Fig. 8 shows the dependence of the flux, i.e. the counts per unit time of particles flowing from one cell to the other, on time. The red (blue) dots in Fig. 8 show the flux from cell 1(2) to cell 2(1) along time. In the initial few seconds, the flow rates from cell 1(2) to cell 2(1) are quickly decreasing at about the same speed. Numbers of particles in two cells are almost equal, and the flux from one cell to the other is nearly equal to the flux of the opposite direction, namely the net flux is nearly zero. At some instant if cell 2 happens to have more particles flowing out than flowing in, less particles will be in cell 2 and less inelastic collisions will cause particles in cell 2 move faster on average than particles in cell 1. There will be a positive feedback to induce even more particles flow out from cell 2 to cell 1 until most of the particles are in cell 1 and almost no more particles can flow from cell 2 to cell 1. At this time the flux from either cell will be the same again, and the net flux will be zero again.

Next, we take the advantage of computer simulation to study the effect of total number of particles in the segregation conditions under no gravity.

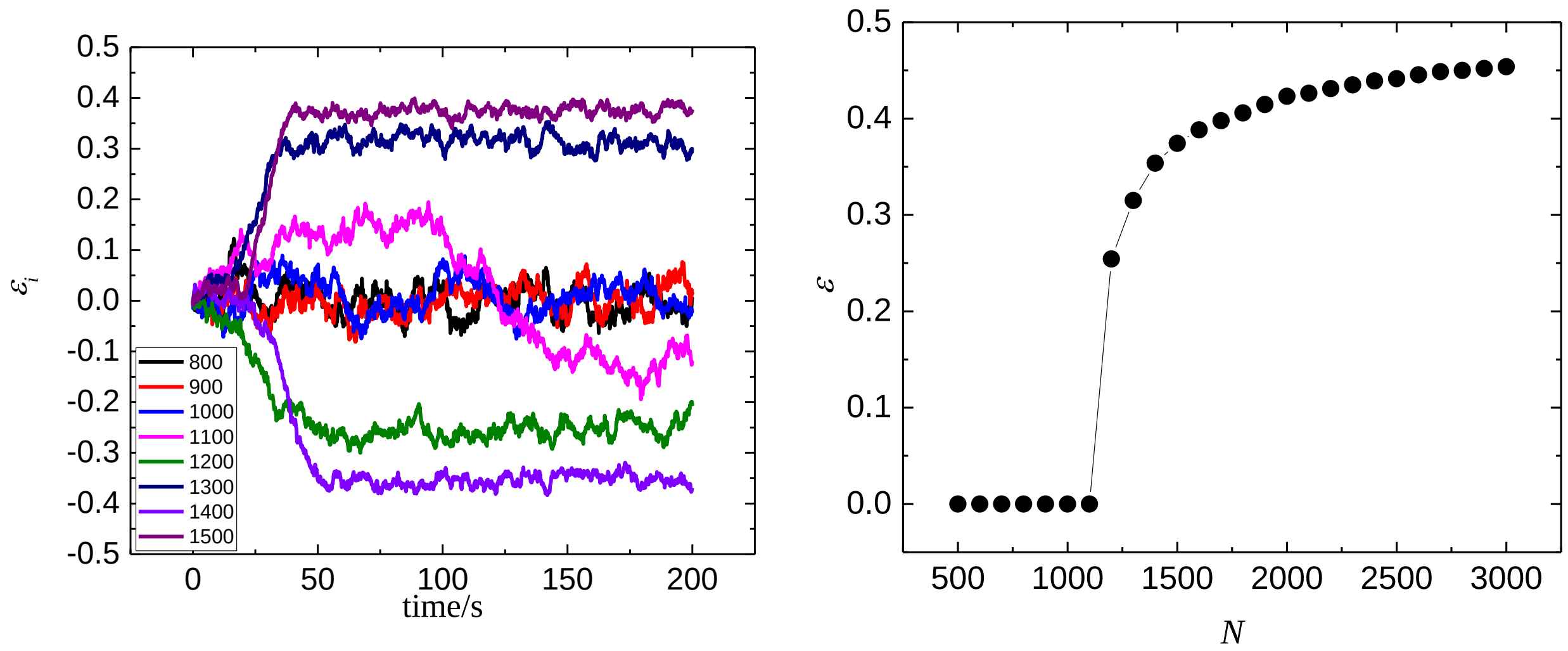

(a) (b)

Fig. 9. (a) Variation of the asymmetry parameters $\varepsilon$ under different $N$ as a function of time; (b) Variation of the asymmetry parameter $\varepsilon$ in the steady state as a function of $N$.

Next we change the total number N from 800 to 1500. Fig. 9(a) gives us the results how $\varepsilon_i(t)$ changes with N. When the total number $N$ is less than or equal to 1100, the asymmetry $\varepsilon$ fluctuate around 0. Segregation does not occur. Only when $N$ exceeds a critical value (here the critical value is about 1100) the segregation occurs. Fig. 9(b) shows that the asymmetry parameter $\varepsilon$ grows from 0 and approaches 0.5 with increasing $N$.

### 4.4. Effect of excitation acceleration on the segregation

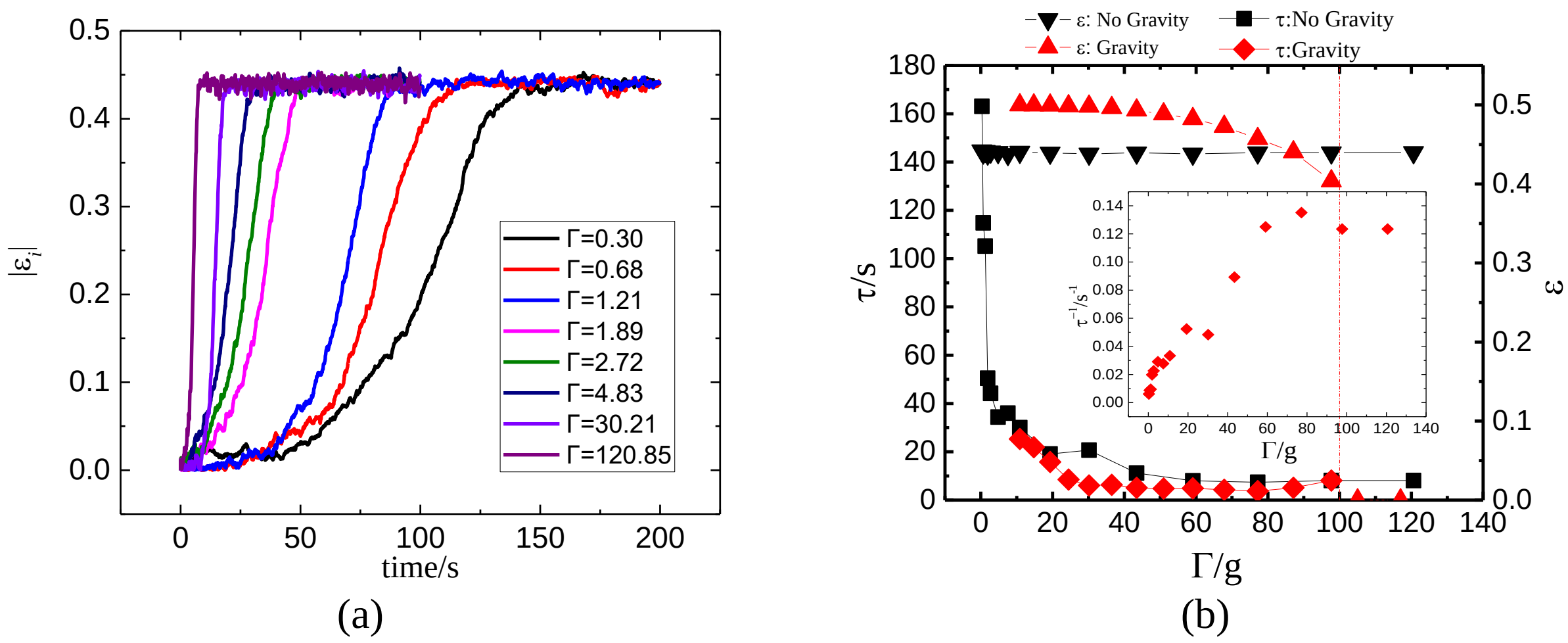

Fig. 10. (a) The asymmetry parameter $|\varepsilon|$ reaches the steady state after waiting time $\tau$ ; (b) Effects of the vibration intensity on the asymmetry parameter $\varepsilon$ and the waiting time $\tau$ under gravity and no gravity.

Fig. 10(a) gives the simulation results of the asymmetry parameter $|\varepsilon_i|$ changing with time under different Γ. For higher acceleration Γ, $|\varepsilon_i|$ reaches steady state value $\varepsilon$ faster.

Since in microgravity experiments the operational time is normally limited, the waiting time $\tau$ is important to know in advance, which tells us how long to wait for seeing the effect. Shown in Fig. 10 (b), the rate $1/\tau$ seems linearly proportional to Γ at low Γ. Only when Γ is high enough the rate $1/\tau$ reaches a maximum value. For the Γ value available in our microgravity experiment (normally less than 1g), Fig. 10(b) tells us the time duration $\tau$ shall be of the order of 100 seconds. This result helps us in determining the experimental design.

In Fig. 10 (b), it shows that the waiting time $\tau$ does not go to zero, there is a minimum time required for particles to go to the steady segregation state no matter how strong the vibration intensity goes. In our case the value is of the order of 10 seconds.

The above numerical results are quite different with the experiments in gravity. With gravity particles act like gas molecules, and distribute in both cells. No segregation of the granular gas is observed at high Γ when particle random collisions dominate. When Γ is below a certain threshold, no segregation is observed either as the gravitational force dominates. With no gravity, the segregation $\varepsilon$ becomes a constant and does not depend on Γ, as seen in Fig. 10(b), but the waiting time $\tau$ depends strongly on Γ.

### 4.5. Effect of the position of the opening on the segregation

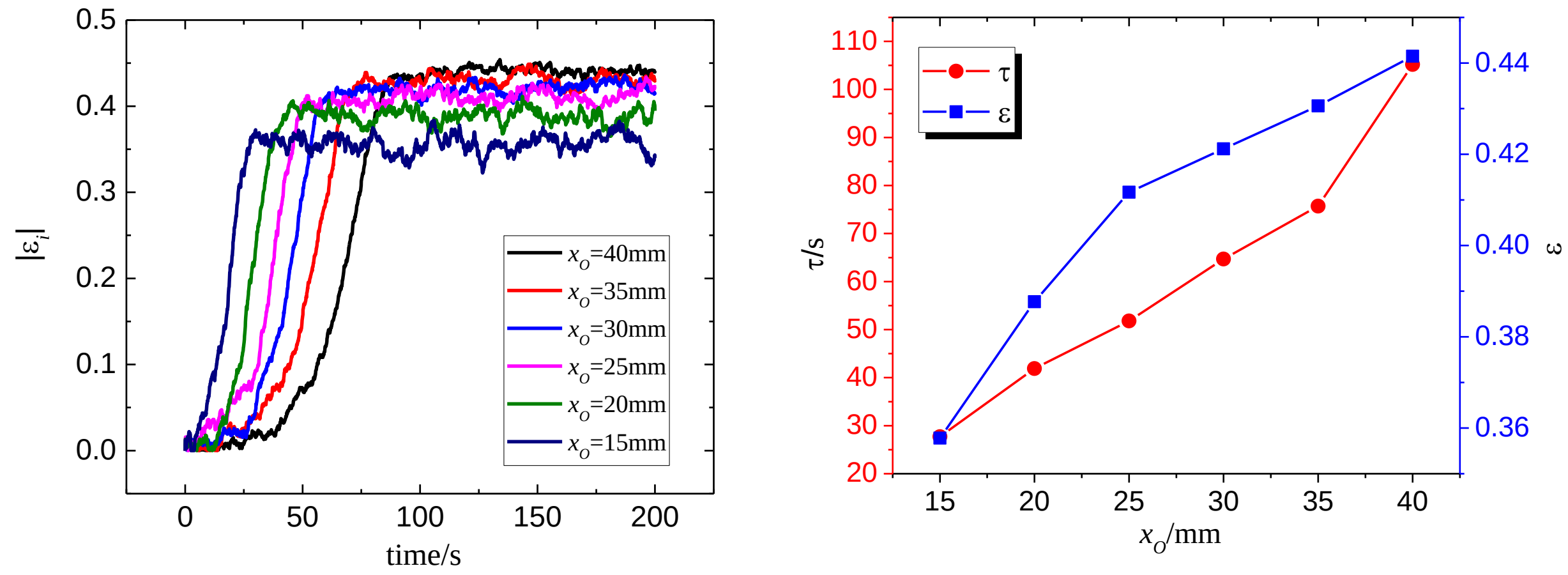

(a) (b)

Fig. 11. (a) Variation of the asymmetry parameter $|\varepsilon_i|$ as a function of time; (b) The waiting time $\tau$ (red triangular dots) and the asymmetry parameter $\varepsilon$ (blue dots) as a function of window position.

The effect of the window position is investigated by changing the piston location. One shall notice that when changing the piston location the length (volume) of the cell also changes. The window connecting the two compartments is located at a position $x_l$ from the position of the left piston. The right piston is accelerated at 4.83 $g$ (where $f_r$ = 20Hz and $A_r$ = 3mm). The effects on both the waiting time $\tau$ and the asymmetry parameter $\varepsilon$ are shown in Fig. 11. The segregation efficiency is better when the window is closer to the cool-end window. As is shown in Fig. 11 (b), the value of $\varepsilon$ changes from 0.44 to 0.36 when $x_l$ changes from 10 mm to 35 mm. It means the segregation is more efficient when the window is closer to the cool-end. However, to reach the steady segregation we need to wait longer time when the window is closer to the cool-end as the time $\tau$ increases, as is seen in Fig. 11 (b).

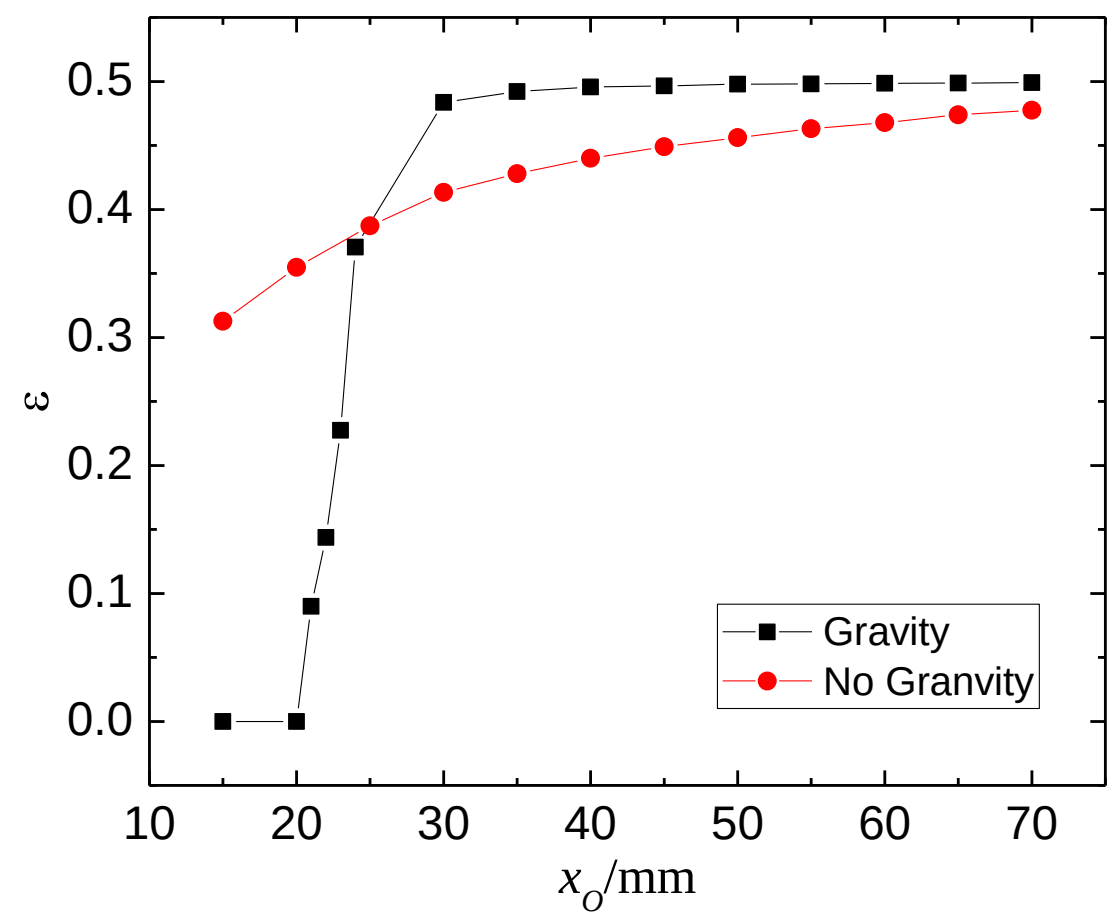

Fig. 12. The asymmetry parameter $\varepsilon$ as a function of opening position with and without gravity.

The previous researches show that the window's position has influence on the occurrence of the segregation under gravity, namely that only when the window's position is closer enough to the bottom, the segregation can appear. Our numerical results in Fig. 12 also show the same result. When gravity does not play a role, the position effect to the segregation becomes weaker.

## 5. Conclusion

Our simulation shows granular segregation is achievable at zero gravity, although the condition can be different from the case under gravity. With gravity, the segregation depends on Γ and can be divided into three regimes: one at low Γ when the gravitational force is dominant that no segregation appears and $\varepsilon$ is zero; the second regime when Γ is at an intermediate value that segregation appears $\varepsilon$ becomes non-zero; the third regime when Γ is high enough that random particle collisions dominate and $\varepsilon$ becomes zero again. Under zero gravity, segregation $\varepsilon$ does not depend on Γ. It is a constant and is determined by the geometry of the cell. A minimum excitation time is necessary to observe the phenomenon with or without gravity. This time $\tau$, however, depends strongly on the value Γ. With the affordable acceleration Γ in satellite condition, the excitation time is on the order of a few minutes. This tells us why one shall not expect segregation in drop tower and parabolic-flight experiments[45] since they offer microgravity for durations from 3 to 22 seconds only.